\documentclass[aps,pra,twocolumn,nofootinbib,superscriptaddress,10pt]{revtex4-1}

\usepackage{natbib}
\usepackage{amsmath,amssymb,bm}
\usepackage{graphicx}
\usepackage{dsfont}

\newcommand{\ket}[1]{|{#1}\rangle}
\newcommand{\bra}[1]{\langle{#1}|}

\begin{document}

\title{Quantum channel correction with twisted light using compressive sensing}

\author{Chemist M. Mabena}
\affiliation{CSIR National Laser Centre, P.O. Box 395, Pretoria 0001, South Africa}
\affiliation{School of Physics, University of the Witwatersrand, Johannesburg 2000, South Africa}

\author{Filippus S. Roux}
\email{froux@nmisa.org}
\affiliation{National Metrology Institute of South Africa, Meiring Naud{\'e} Road, Brummeria, Pretoria, South Africa}
\affiliation{School of Physics, University of the Witwatersrand, Johannesburg 2000, South Africa}

\begin{abstract}
Compressive sensing is used to perform high-dimensional quantum channel estimation with classical light. As an example, we perform a numerical simulation for the case of a three-dimensional classically non-separable state that is propagated through atmospheric turbulence. Using singular value thresholding algorithm based compressive sensing, we determine the channel matrix, which we subsequently use to correct for the atmospheric turbulence induced distortions. As a measure of the success of the procedure, we calculate the fidelity and the trace distance of the corrected density matrix with respect to the input state, and compare the results with those of the density matrix for the uncorrected state. Furthermore, we quantify the amount of classical non-separability in the density matrix of the corrected state by calculating its negativity. The results show that compressive sensing could contribute in the development and implementation of free-space quantum and optical communication systems.
\end{abstract}

\maketitle

\section{Introduction}

Quantum entanglement in higher dimensions is a property that can enhance communication security using quantum key distribution \cite{gibson, barreiro, cryptotwist}. The spatial degrees of freedom of single-photon fields allow infinitely many modes, such as the orbital angular momentum (OAM) modes \cite{allen}, which can be used to design and prepare such higher dimensional quantum systems. Optical fields of OAM modes have an azimuthal phase dependence given by $\exp(i\ell \phi)$, with $\phi$ the azimuthal angle. They carry quantized OAM of $\ell \hbar$ per photon. Photons that have the above described phase dependence and carry OAM are commonly referred to as {\em twisted photons}.

Free-space optical and quantum communication systems based on the OAM states of light, are adversely affected by turbulence in the atmosphere, which leads to a distortion of the OAM states \cite{paterson, scintbook, numipe}. This distortion further results in the decay of quantum entanglement, which has been studied both theoretically \cite{sr, qturb4, qturb3,ipe,leonhard,bruenner} and experimentally \cite{pors,malik,oamturb,qkdturb,qutrit}. Our understanding of the behavior of OAM-based quantum systems indicates that the potential offered by the OAM states of light for free-space communication requires efficient methods to mitigate against the adverse effects of atmospheric turbulence.

The standard method of characterizing a quantum channel is called standard quantum process tomography (SQPT) \cite{nc}. The characterization and determination of quantum processes are crucial tasks necessary for the implementation of quantum communication and information processing systems, enabling the design of mitigation strategies against noise and other distortions that may negatively affect a quantum system. However, SQPT is a resource intensive process that scales dramatically with the size of the quantum system \cite{nc,mohseni}. The number of resources required for SQPT is generally $O(N^4)$, where $N$ is the dimension of the Hilbert space.

Scintillation, caused by atmospheric turbulence, leads to the distortion of input OAM modes, distributing the power into other modes \cite{paterson,tyler}, thus increasing the dimensionality of the photonic quantum systems when expanded in terms of OAM modes \cite{anguita}. The increase in dimensionality caused by scintillation implies that the application of SQPT would not be favorable for characterizing a turbulent atmospheric channel. The mechanism that produces scintillation is a pure phase modulation caused by the fluctuating refractive index of the atmosphere. It does not produce a loss prior to the measurement stage. Although the atmosphere does have additional loss mechanisms such as absorption and scattering, we do not include them in our current investigation, for reasons explained below.

For successful implementation of quantum communication systems with twisted photons, a different method of determining and characterizing the turbulent quantum channel is needed. This is where we employ {\em compressive sensing} \cite{compsens1, compsens2, compsens3} --- a data processing method that provides an efficient mechanism for the recovery of unknown signals from only a fraction of the required measurements. A generalization of this method to matrices is called matrix completion \cite{matcompl}. In traditional compressive sensing the signal is required to have certain properties such as sparseness in the appropriate basis. In the matrix generalization case, the matrices must also obey certain criteria. For instance, the method is more likely to be successful for low rank matrices. Fortunately, this is the case for a single realization of atmospheric turbulence --- a snapshot of the atmospheric conditions along the entire path of the channel as experienced by a single light pulse. Scintillation is treated as a unitary process that preserves the purity of quantum states.

Gross \textit{et al.} \cite{gross} have established a method based on a random selection of Pauli measurements for efficient reconstruction of an unknown quantum state. They showed that their method can reconstruct a rank $r$ unknown density matrix with only $O(rN\log^2N)$ measurements, in contrast to the $O(N^2)$ measurements for the standard method. In another study \cite{tonolini}, a high-dimensional entangled state was reconstructed from a significantly smaller number of measurements, using a related approach based on compressive sensing. While this method for the reconstruction of signals from an underdetermined system of equations is very popular in signal and image processing applications, it has now attracted interest and become topical in quantum information science related applications as well \cite{compsens4,compsens5, compsens6, compsens7,compsens8,gross}.

The aim of this work is to show with the aid of numerical simulations that compressive sensing is a viable technique for channel estimation in the specific case of the transmission of twisted light through atmospheric turbulence. Moreover, inspired by a recent study on the characterization of a quantum channel with classical light \cite{ndagano,chanup}, we developed a more efficient protocol for characterizing turbulent channels by combining it with compressive sensing. In turn, it leads to an efficient scheme for the correction of twisted photons after passing through atmospheric turbulence.

The proposed scheme does not suffer any detrimental effects due to loss, because it is based on classical light (a bright coherent state). In the context of quantum states, a probability distribution of different losses would cause the quantum state to become mixed. However, the proposed scheme allows one to make the pulse shorter than the time scale given by the Greenwood frequency\cite{greenwood} of the turbulent medium. As a result, the accumulated loss that the pulse experience does not represent a probability distribution. The pulse only experiences a single constant accumulated loss. Hence, no mixing takes place and the state remains pure. For this reason, we do not include any loss mechanisms in our current investigation.

The compressive sensing part requires relatively few random measurements, while the use of classical light removes the intrinsic limitation of quantum mechanics that requires repeated measurements on an ensemble. Also, since classical light is extremely bright compared to a few discrete photons, we can perform the different random measurements at the same time, thus leading to a significant speed up in time. The measurements on the classical light upon propagating through atmospheric turbulence will be performed using sum-frequency generation as previously discussed in Ref. \cite{chanup}. Measurements on the output state are made by optically combining the output state with a tailored measurement state through a non-linear crystal, the measurement state selects out a specific component in the output state. This entire process leads to a photon detection which represents a measurement. The compressive sensing model used in this work is based on the singular value thresholding algorithm \cite{svthres} that has been modified in a way similar to the one used in Ref. \cite{tonolini}.

The outline of this contribution is as follows. In Sec.\ref{model} we present the model for this work.  The numerical simulation method is discussed briefly in Sec.~\ref{NS}. Section~\ref{Res} is based on the results of this work. Finally, the conclusions are given in Sec.~\ref{conclusions}.

\section{Model}
\label{model}

\subsection{Channel matrix}

The Choi-Jamiolkowski isomorphism \cite{chandual} establishes a correspondence between a completely positive trace-preserving quantum map $\Lambda$ and a quantum state $\rho$ by
\begin{eqnarray}
\rho_{\Lambda} = (\Lambda_{A} \otimes \mathds{1}_{B})(\ket{\psi} \bra{\psi}), \label{choi}
\end{eqnarray}
where
\begin{eqnarray}
\ket{\psi} = \frac{1}{\sqrt{N}}\sum_{n = 1}^{N}\ket{n}_{A}\otimes\ket{n}_{B}, \label{st1}
\end{eqnarray}
is a maximally entangled state, $\mathds{1}$ is the identity operator for subsystem B, and $N$ is the dimension of the Hilbert space. This isomorphism means that the identification and characterization of the quantum channel is tantamount to performing a quantum state tomography.

Here, the partites of the state in Eq.~(\ref{st1}) are represented by the spatial modes and the wavelengths of a classical optical field. A perfect correlation between these degrees of freedom (maximal nonseparability) gives us an exact analogy with a maximally entangled quantum state. So, the typical input state for our consideration reads \cite{chanup}
\begin{equation}
\ket{\psi} = \frac{1}{\sqrt{2M + 1}}\sum_{\ell = - M}^{M} \ket{\ell,0}_{A}\ket{\lambda_\ell}_{B} ,
\label{choiIso}
\end{equation}
where $\ket{\ell,0}$ represents a Laguerre-Gaussian (LG) mode with azimuthal index $\ell$ and radial index $p=0$; $\lambda_\ell$ is the wavelength of the corresponding LG mode; and $M$ is an integer representing the maximum OAM. In general, one can have an arbitrary radial index, so that the LG mode would be $\ket{\ell,p}$. However, for the moment, we assume correlation between the wavelength and the azimuthal index and therefore we set $p=0$.

Transmission through the atmosphere causes the OAM modes to scatter into other modes. The atmospheric turbulence only affects the spatial degree of freedom and leaves the wavelength unaffected. Therefore, after propagating through the atmosphere, a given input OAM mode becomes
\begin{equation}
\ket{\ell,p} \rightarrow \sum_{{\ell}',p'} \ket{\ell',p'} \Lambda_{{\ell},p}^{{\ell}',p'} ,\
\end{equation}
where $\ket{\ell',p'}$ is an LG mode with azimuthal index $\ell'$ and radial index $p'$; $\Lambda_{{\ell},p}^{{\ell}',p'} $ is the tensor representation of the atmospheric turbulence Kraus operator. It is this tensor that represents the effect of the atmosphere. We need to characterize it to mitigate against its effect.

To simplify notation, we'll index the input and output modes by single integers. Hence, the scattering process is represented by
\begin{equation}
\ket{n} \rightarrow \sum_m \ket{m} \Lambda_{n}^{m} .\
\end{equation}
where the Kraus operation is now represented by an $N\times M$ matrix. The Krauss operator can thus be represented, either in terms of the tensor or in terms of a matrix
\begin{align}
\Lambda = & \sum_{{\ell},p,{\ell}',p'} \ket{\ell',p'} \Lambda_{{\ell},p}^{{\ell}',p'} \bra{\ell,p} \nonumber \\
= & \sum_{m,n} \ket{m} \Lambda_{n}^{m} \bra{n} .
\label{krausop}
\end{align}

In the most general case, the matrix $\Lambda_{n}^{m}$ is rectangular $M\neq N$. The input state is usually defined in terms of a finite number of modes. So, the dimensionality of the input Hilbert space is finite and determines the number of columns $N$ of the matrix. On the other hand, the number of rows $M$ is determined by the crosstalk induced by the atmospheric turbulence. Since there are an infinite number of spatial modes and since the scintillation process can potentially scatter the input modes into any combination of output modes, one can expect that $M\gg N$ and that $M\rightarrow\infty$. However, the scattering is not uniform --- the dominant scattering tends to produce modes lying close to the input modes. One can therefore truncate the output space to a finite number of dimensions, depending on the accuracy that is required.

\subsection{\label{compsense}Compressive sensing}

Here we briefly review the compressing sensing procedure that we used for our work. For this purpose, we follow the procedure of Tonolini {\em et al.} \cite{tonolini}.

Using the matrix completion method \cite{matcompl}, one can recover a low-rank matrix when some of the elements of the matrix are unknown. Instead of recovering the density matrix from a sample of its elements $\rho_{i,j}$, we recover the full matrix that represents the states in terms of a sample of the results of measurements made on the system.

In the most general case, the density matrix can be decomposed in terms of the Bloch representation
\begin{equation}
\rho = \sum^{N^2}_{i=1} \alpha_{i} \tau_{i} ,
\end{equation}
where $\alpha_{i}$ are the elements of the Bloch vector, $N$ represents the dimensionality of the state vector and $\tau_i$ denotes the generalized Gell-Mann matrices (GGMs), including the identity matrix. The GGMs reduce to the Pauli matrices in the case where $N=2$. In this formalism, the GGMs form a convenient measurement basis for the characterization of the state. To determine the state through a full tomography, one must perform measurements that reveal $N^2$ real parameters $\alpha_i$. The Bloch vector elements are the expectation values given by the trace
\begin{equation}
\alpha_i = \text{tr}\left\{\tau_i \rho \right\} .
\label{chp2}
\end{equation}

In the implementation of the modified matrix completion problem \cite{gross}, we consider an under-sampled set of measurements ($m \ll N^2$) chosen at random --- i.e., we consider a situation where there is only a subset of the total possible measurements $\alpha_i$. The optimization problem is thus described as follows: Minimize $||\rho_\text{r}||_\text{tr}$ such that
\begin{align}
\begin{split}
\text{tr}\{\rho_\text{r}\} & = 1 , \\
\rho_\text{r} & = \rho_{\text{r}}^{\dagger} , \\
\text{tr}\{\rho_{\text{r}}\tau_{i}\} & = \alpha_i ~~ \text{for} ~~ i = 1...m ,
\end{split}\
\end{align}
where $\rho_\text{r}$ is the to-be-recovered density matrix, and $||\cdot||_\text{tr}$ is the trace norm of the matrix, given by
\begin{equation}
||\rho_\text{r}||_\text{tr} = \text{tr}\left\{\sqrt{\rho_\text{r}\rho_\text{r}^{\dagger}}\right\} .
\end{equation}

The compressive sensing algorithm is based on the singular value thresholding (SVT) algorithm \cite{svthres}. However, we modify the algorithm to take advantage of known properties of the state that we intend to recover \cite{tonolini}. Essentially, in applying the SVT algorithm we perform an eigenvalue decomposition of the density matrix
\begin{equation}
\rho_\text{r} = \sum_{j}\ket{\phi_{j}}\sigma_{j}\bra{\phi_{j}},
\end{equation}
where $\ket{\phi_{j}}$ are the eigenvectors of $\rho_\text{r}$ and $\sigma_{j}$ are the corresponding eigenvalues. Given the above decomposition, we apply the thresholding operator on the eigenvalues, by selecting the eigenvalues above a certain threshold $\epsilon_0$ that we fixed a priori. Ensuring that the eigenvalues of the density matrix are real, we force the density matrix to be Hermitian. The normalization of the density matrix is achieved by dividing the resultant matrix by its trace.

The algorithm uses a guess matrix as a starting point for the optimization process. The most crucial part in setting up the guess matrix is choosing its dimensions, because the dimensions of the matrix must be representative of the atmospheric turbulence conditions. The matrix must be big enough to accommodate most of the non-zero elements in the OAM modal spectrum. Once we have determined the dimensions of the guess matrix, we can apply the algorithm.

After performing the thresholding on the density matrix and normalizing it, we obtain a density matrix that may represent a real physical system, but it no longer corresponds to the density matrix whose measurements gave the correct measurement results $\{\alpha_i\}$ with respect to the GGMs $\tau_i$ for $i=1...m$ \cite{tonolini}. To understand the reason for this issue, we use a geometrical perspective, represented in terms of hyperplanes. For this purpose, we re-express Eq.~(\ref{chp2}) in vector notation:
\begin{equation}
M \bar{\rho} = \bar{\alpha} ,
\label{chp2Re}
\end{equation}
where $\bar{\rho}$ is the vectorized density matrix, $M$ is a matrix with rows representing the vectorized GGM, and $\bar{\alpha}$ is the vector of the measurement results. In this representation, each vectorized GGM $\bar{\tau}_i$ and its corresponding measurement result ${\alpha}_i$ can be associated with a hyperplane in an $N^2$ dimensional space. As such, the compressive sensing approach attempts to solve an under-determined linear system of equations. The solution should be a single point common to all the $m$ hyperplanes. However, the procedure explained above does not produce a point that lies in the linear space of Eq.~(\ref{chp2Re}):
\begin{equation}
M \bar{\rho}_\text{r} \neq \bar{\alpha}.
\label{chp2Re2}
\end{equation}

To resolve this issue, we project the density matrix back into the linear space, determined by ${\tau}_i$ and their corresponding results $\alpha_i$. In the process, we modify the density matrix $\rho_\text{r}$ to give the correct measurement results. The projection is done stepwise for each value of $\alpha_i$. The projection starts by defining a vector $\bar{v}_i$ normal to the hyperplane $\bar{\tau}_i \bar{\rho} = \alpha_i$, and with a magnitude
\begin{equation}
|\bar{v}_i| = \alpha_i - \hat{v}\cdot\bar{\rho}_{i-1},
\end{equation}
where $\hat{v}$ is the unit vector that is normal to the hyperplane and $\bar{\rho}_{i-1}$ is the density matric obtained in the previous iteration. The new density matrix becomes
\begin{equation}
\bar{\rho}_{i} = \bar{\rho}_{i-1} + \bar{v}_i.
\end{equation}
The projection process is performed for all $m$ measurements. At the end of this process, we start again and perform the procedure for Hermiticity and trace normalization. In other words, we perform the thresholding operation and further recompose the matrix according to the conditions in the optimization problem statement. All these steps are performed iteratively until the norm of the difference between the density matrices of two consecutive iterations lies within a predefined tolerance.

\subsection{Channel correction}
\label{chanCor}

Here follows the main contribution of our work. It involves the generation of the Kraus operator matrix from the estimated density matrix, obtained from the compressive sense procedure described above.

The full Kraus operator matrix can be obtained from the density matrix of the output state. This step is made possible by the fact that, despite the randomness of the medium, propagation through a turbulent atmosphere is a unitary process. Therefore, the output quantum state after transmission through a single realization of atmospheric turbulence is always a pure state, provided that the input state was a pure state. We also assume that the truncation of the output space to a finite dimension does not affect the unitarity of the process significantly, provided that we use a large enough number for the output dimension. As a result, we consider the reconstructed Kraus operator as a unitary operator, so that its Hermitian adjoint represents the inverse process.

After the Kraus operator $\Lambda$ has been reconstructed, using the compressive sensing methods described above, we perform the correction process as follows
\begin{align}
\ket{\Psi}_{\text{in}} = & \left( {\Lambda}_{A}\otimes\mathds{1}_{B} \right)^{\dagger}\ket{\Psi}_{\text{out}} \nonumber\\
= & \left( \sum_{m,n} \ket{m} \Lambda_{n}^{m} \bra{n} \otimes\mathds{1}_{B} \right)^{\dagger}\ket{\Psi}_{\text{out}} .
\end{align}
The validity of the unitary assumption is assessed by the quality of the correction, as discussed below.

\section{\label{NS}Numerical simulation and computations}

We performed numerical simulations to test the proposed compressive sensing-based scheme for quantum channel estimation and correction. For this purpose, we consider a three-dimensional bipartite input state (qutrit). The two degrees of freedom that represents the two partites are the spatial mode (OAM mode) and the wavelength. The input state can be expressed as
\begin{equation}
\ket{\Psi} = \frac{1}{\sqrt{3}}\left[ \ket{\ell,0}\ket{\lambda_\ell} + \ket{0,0}\ket{\lambda_0} +  \ket{\bar{\ell},0}\ket{\lambda_{\bar{\ell}}} \right].
\label{inputState}
\end{equation}
In the simulation, the input state is represented by three $n\times n$ sampled functions for the three spatial modes, where $n=1024$. Note that Eq. (\ref{inputState}) represents a classical field that is nonseparable (as opposed to being entangled), expressed in terms of Dirac notation.

Each of the modes are separately propagated in the simulation process. The propagation of paraxial optical fields in atmospheric turbulence is described by the stochastic parabolic equation
\begin{equation}
\partial_z{f(\mathbf{r})} = \frac{i}{2 k_0} \nabla_\perp f (\mathbf{r}) - i k_0 \delta n(\mathbf{r})f\left(\mathbf{r} \right) ,
\label{parB}
\end{equation}
where $\nabla_\perp = \partial^2_x + \partial^2_y$, $\delta n(\mathbf{r})$ is the refractive index fluctuation of the atmosphere ($n=1+\delta n$), $k_0 = 2\pi/\lambda$ is the wave number, and $\lambda$ is the wavelength of the optical field.

In weak scintillation, the only effect of the atmospheric turbulence is a phase perturbation on the optical field. It means that propagation through the atmosphere under weak scintillation conditions can be represented by two steps. The first step is a random phase modulation of the input optical field, which represents the perturbation of the field and leads to refraction. The subsequent step is free-space propagation (without turbulence) over the full propagation distance.

For arbitrary scintillation conditions, one can still use these two steps to simulate propagation through turbulence. However, one would repeat the two steps multiple times, each time propagating over a short enough distance to ensure weak scintillation conditions for that step. During each step, the optical field is modulated by a different random phase screen. Therefore, to simulate the propagation of an optical field through a turbulent atmosphere we use a split-step method \cite{anguita}, in which these two steps are repeated several times.

The generation of the random phase screen entails transforming a 2D-array of random complex numbers that have zero mean and unit variance into an array that has the same statistics as the atmospheric turbulence. This process is also known as filtering Gaussian noise and is given by \cite{mf1,knepp}
\begin{equation}
\theta(\mathbf{R}) = \left(2 \pi k_0^2\Delta z  \right)^{1/2}\mathcal{F}^{-1}\left\{ \chi(\mathbf{K}) \left[ \frac{\Phi_{n}(\mathbf{K})}{\Delta^2_{k}} \right]^{{1}/{2}}\right\},
\label{PhScr}
\end{equation}
where $\Delta z$ is the partitioned propagation distance between two consecutive phase screens, $\mathcal{F}^{-1}\{\cdot\}$ denotes the inverse Fourier transform, $\Phi_{n}(\mathbf{K})$ is the Kolmogorov power spectral density (PSD) for the refractive index, $\Delta_{k}$ is the grid-spacing in the spatial frequency domain, and $\mathbf{K}=(k_x,k_y)$ is the transverse wavevector. The normally distributed complex random function $\chi(\mathbf{K})$ has zero mean and is $\delta$-correlated, such that
\begin{equation}
\langle \chi(\mathbf{K}_1)\chi^{*}(\mathbf{K}_2)\rangle = \left(2 \pi \Delta_{k}\right)^2 \delta(\mathbf{K}_1 - \mathbf{K}_2) ,
\end{equation}
where the angled brackets $\langle\cdot\rangle$ denote an ensemble average. The random phase function generated by Eq.~(\ref{PhScr}) is complex [unless $\chi^{*}(\mathbf{K})=\chi(-\mathbf{K})$]. Hence, a single application of Eq.~(\ref{PhScr}) produces two independent random phase screens --- the real and imaginary parts of the complex phase function.

It is important to point out that the Fourier calculation in Eq.~(\ref{PhScr}) does not take into account the effect of large eddies, which are excluded due to the discrete grid samples in the Fourier domain. As a result, the statistics obtained from such phase screens do not represent the Kolmogorov structure function correctly. One way to improve the accuracy is to add sub-harmonics \cite{dainty} to the phase function generated by the FFT method in Eq.~(\ref{PhScr}). The sub-harmonic phase function is given by
\begin{align}
\theta_{SH}(j\Delta x, l\Delta y) = & \sum_{n = 1}^{N_s} \sum_{p, q = -1}^{ 1} \left[ a(p,q,n) + ib(p, q, n)\right] \nonumber \\
& \times \exp\left[2 \pi i \left(  \frac{jp}{3^n N_x} + \frac{lq}{3^n N_y} \right) \right] .
\end{align}
The variance of the randomly generated functions $a$ and $b$ is given as,
\begin{align}
\langle a^2(p,q,n)\rangle = & \langle b^2(p,q,n)\rangle \nonumber \\
 = & \Delta {p_n} \Delta {q_n} \Phi_{\theta}(p \Delta {p_n} ,q \Delta {q_n}) ,
\end{align}
where, $\Delta{p_n}=\Delta p/3^n$, $\Delta{q_n}=\Delta q/3^n$, $N_x$ and $N_y$ are the numbers of points in the $x$ and $y$ directions, respectively, and $N_s$ is the number of sub-harmonics.

In the simulation, we propagate the input state through the turbulent medium by performing the split-step process on each of the three input modes, with their associated wavelengths. The two-step process is iterated several times to perform a multi-phase screen propagation of the input state through a realization of the turbulent medium. The complete propagation is then done for several realizations.

The parameters that are used for the propagation process are: input beam waist radius $w_0=0.1$~m; propagation distance $z=2z_R$, where $z_R=\pi w_0^2/\lambda$ is the Rayleigh range; and turbulence strength (refractive index structure constant) $C_n^2=1\times 10^{-16}~\text{m}^{-2/3}$. The wavelengths of the three modes are different, however, in practical setups one can always make these differences to be very small. For the purpose of the simulations, the following wavelengths were used for the different modes: $\lambda_1 = 1.0~\mu \text{m}$, $\lambda_2 = 1.020~\mu \text{m}$, and $\lambda_3 = 1.040~\mu \text{m}$.

The turbulent medium causes crosstalk, transferring power to numerous higher order LG modes. Therefore, the dimensions of the Kraus operator could be very large. The strength of the crosstalk and the dimensionality of the output state depends on various parameters, including the strength of turbulence and the distance of propagation. These parameters combine to determine the strength of the scintillation --- the extent of the distortion imparted by the medium on the state.

\begin{figure}[ht]
\centering
\includegraphics[width = 0.7\linewidth]{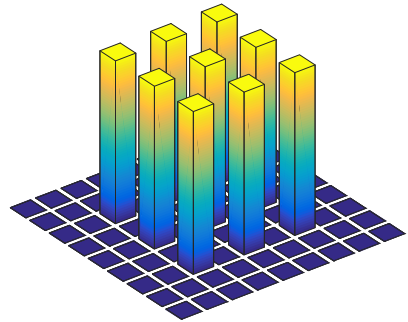}
\caption{Graphical representation of the input state density matrix.}
\label{inputM}
\end{figure}

The dimension of the Hilbert space of the input state is $N_\text{in} = N_{A}\cdot N_{B}$. For the state described in Eq.~(\ref{inputState}), we have $N_A=N_B=3$, which implies that $N_\text{in}=9$. The corresponding density matrix has $81$ elements and is depicted graphically in Fig.~\ref{inputM}.  A unitary process does not change the minimum number of modes required to represent the state. However, the nature of these modes is unknown due to the lack of a priori knowledge of the unitary process associated with a single realization of the turbulent medium. As a result, the output state must be represented in terms of some nominal modal basis.

Here, we use the LG modes as our chosen output modal basis. While the input state has only three-dimensions per degree of freedom, the dimension of the output state in terms of the LG modes is much larger. The turbulent medium affects the dimension of the one subsystem $N_{A}\rightarrow N_{A}'$ but leaves the that the other $N_{B}$ the same. The dimension of the output density matrix thus becomes $N_\text{out} \times N_\text{out}$, where $N_\text{out} = N_{A}'\cdot N_{B}$.

In our simulation, we include radial indices up to $p = 7$. For each value of the radial indices, we considered azimuthal indices in the range $-14 \leq \ell \leq 15$. It implies that the output dimension in subsystem $A$ becomes $N_A'=210$, so that the total output dimension becomes $N_\text{out} = 630$. This number is required to represent the output beam profiles with adequate fidelity for the parameters that we used in the simulation. The output density matrix thus has almost $400~000$ elements. To specify a density matrix obtained in this way, one needs as many measurements.

\begin{figure}[ht]
\centering
\includegraphics[width = 0.8\linewidth]{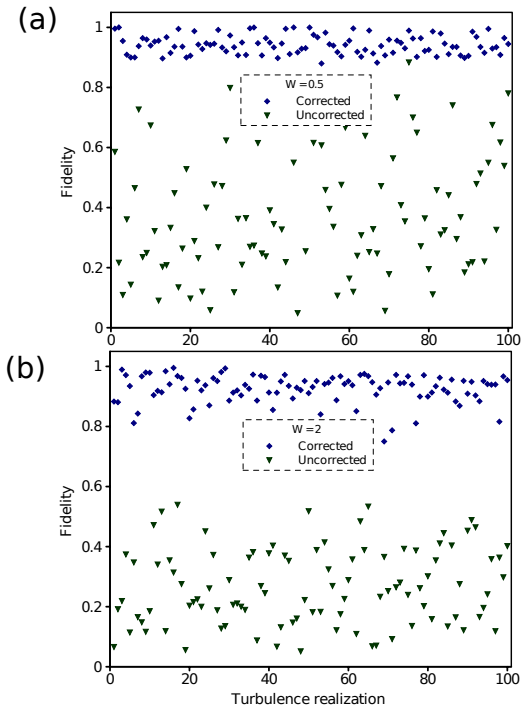}
\caption{Fidelity of the corrected and the uncorrected density matrices for the different atmospheric turbulence realizations and for (a) $W=0.5$ and (b) $W=2$. Triangle markers represent the fidelities for the uncorrected, truncated density matrices. Diamond markers represent the fidelities for the corrected density matrices.}
\label{Fidelity_qtrit}
\end{figure}

However, the state remains pure for each realization of the atmospheric turbulence, because the process is unitary. As a result, the Kraus operator matrix associated with that realization has a low rank. The purity of a state implies that it is represented by a state vector $\ket{\psi_\text{out}}$. The density matrix of such a pure state has rank equal to unity, which allows us to use compressive sensing for estimating the output state and thereby determining the Kraus operator matrix. Using compressive sensing, we can determine the state of the output density matrix to a high level of accuracy, using a significantly smaller number of measurements.

For this work, we use approximately $5\%$ of the total number of measurements (which in this case is roughly 20000) to reconstruct the output density matrix reliably. Using the reconstructed density matrix, we extract the elements of the Kraus operator for the atmospheric turbulence. To test the reliability of the reconstruction we use the Kraus operator to generate the correction matrix and apply it to the output density matrix. The result is then tested for fidelity and trace distance against the input qutrit state Eq.~(\ref{inputState}).

\begin{figure}[ht]
\centering
\includegraphics[width = 0.8\linewidth]{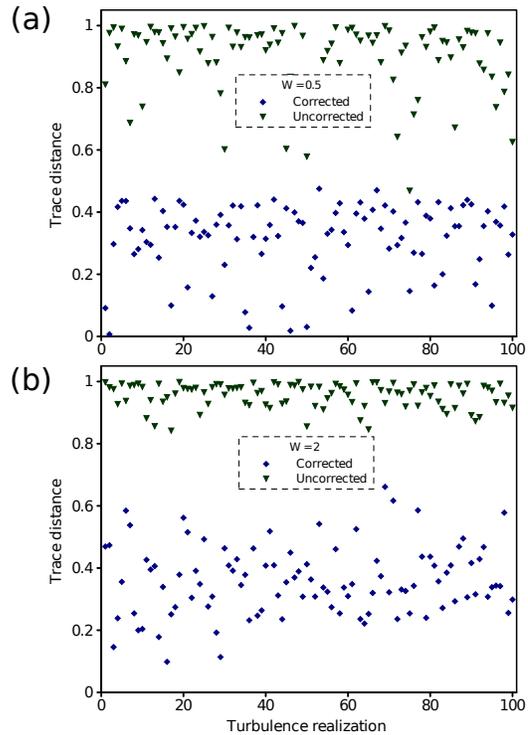}
\caption{Trace distance of the corrected and the uncorrected density matrices for the different atmospheric turbulence realizations and for (a) $W=0.5$ and (b) $W=2$. Triangle markers represent the uncorrected, truncated density matrices. Diamond markers represent the corrected density matrices.}
\label{TraceDist_qtrit}
\end{figure}

Upon application of the compressive sensing algorithm described in Sec.~\ref{compsense}, a density matrix of the output state is obtained. Since a single realization of atmospheric turbulence is a unitary process, the output state is pure. Furthermore, given that the output density matrix is the outer product of the state vector and its adjoint, it follows that a single element of the density matrix is a product of two elements. Hence, by choosing one column or row of the output density matrix and dividing it by the relevant element, the elements of the state vector can be obtained, apart from an overall phase constant. These elements are then rearranged to form the Kraus operator $\Lambda^{m}_{n}$, which is subsequently used to correct the scintillation as described in Sec.~\ref{chanCor}.

\section{Results}
\label{Res}

The numerical simulations of the propagation and correction of OAM states in atmospheric turbulence was performed several times for different realizations of a turbulent medium. The performance of the compressive sensing-based correction scheme is assessed by calculating the fidelity and the trace distance from these results.

\begin{figure}[ht]
\centering
\includegraphics[width = 0.8\linewidth]{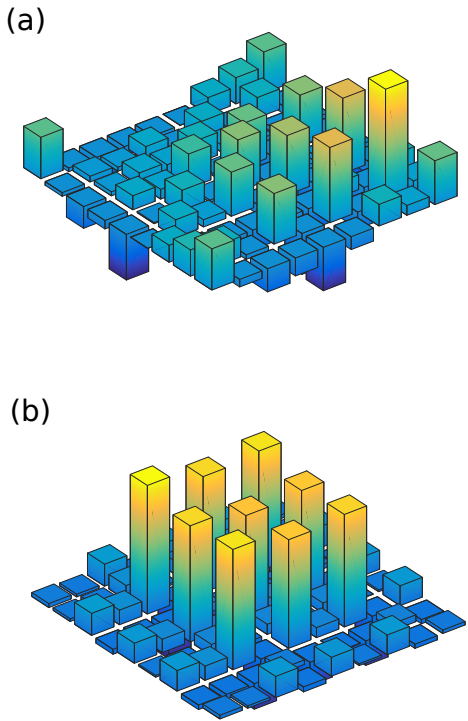}
\caption{Graphic representations of the (a) uncorrected, truncated density matrix, and the (b) corrected density matrix, after passing through the same simulated turbulence represented by the scintillation strength $W=1$.}
\label{randTurb_qtrit}
\end{figure}

The fidelity with respect to the initial maximally nonseparable state in Eq.~(\ref{inputState}) is given by
\begin{equation}
{F}(\rho_\text{c},\ket{\Psi}\bra{\Psi})=\text{tr}\left\{\sqrt{\sqrt{\rho_\text{c}}\ket{\Psi}\bra{\Psi}\sqrt{\rho_\text{c}}}\right \} , \label{fidGeneral}
\end{equation}
where $\ket{\Psi}$ is the input state, and $\rho_\text{c}$ is the corrected density matrix from our compressive sensing algorithm. The results of these calculations are shown in Fig.~\ref{Fidelity_qtrit} for two different scintillation strengths: $W=0.5$ in Fig.~\ref{Fidelity_qtrit}(a) and $W=2$ in Fig.~\ref{Fidelity_qtrit}(b). Here, the scintillation strength is represented by the dimensionless number $W=w_0/r_0$, where $r_0$ is the Fried parameter \cite{fried}, which is given by
\begin{equation}
r_0 = 0.185 \left(\frac{\lambda^2}{C_n^2 z}\right)^{3/5} .
\end{equation}

The trace distance, which is shown in Fig.~\ref{TraceDist_qtrit}, for the same two scintillation strengths, is defined as
\begin{equation}
D(\rho_\text{c},\ket{\Psi}\bra{\Psi}) = \frac{1}{2}\text{tr} \left\{\left| \rho_\text{c} - \ket{\Psi}\bra{\Psi}  \right|\right\},
\end{equation}
where the magnitude $|A|$ of a matrix is given as $|A| = \sqrt{A^\dagger A}$. The data in Figs.~\ref{Fidelity_qtrit} and \ref{TraceDist_qtrit} are plotted against the different realizations of the atmospheric turbulence.

In Fig.~\ref{Fidelity_qtrit}, it is observed that the corrected density matrices are close to the ideal maximally nonseparable input state, in contrast to the uncorrected, truncated density matrices. The mean fidelity for the corrected density matrices over the 100 turbulence realizations in the first case with $W=0.5$ is $0.942 \pm 0.003$, while the uncorrected density matrices give a mean fidelity of $0.38 \pm 0.02$. Here, the error is given as the standard error. For the second case, $W=2$, the mean fidelity of the corrected density matries over 100 turbulence realizations is $0.925 \pm 0.005$ and that of the uncorrected density matrices is $0.265 \pm 0.012$.

A similar conclusion about the success of the correction method can be reached by looking at Fig.~\ref{TraceDist_qtrit}, which shows the trace distance.  The mean trace distance for the corrected density matrices for $W = 0.5$ is $0.31 \pm 0.02$, while that of the uncorrected density matrices is $0.892 \pm 0.011$. For the second scintillation strength, $W=2$, the mean trace distance for the corrected density matrix is $0.359 \pm 0.011$ and for the uncorrected density matrices it is $0.955 \pm 0.004$.

An example of the real parts of the elements of an uncorrected and corrected density matrix is shown in Fig.~\ref{randTurb_qtrit} for an arbitrary turbulence realization. Comparison of the unperturbed density matrix in Fig.~\ref{inputM} with the density matrices in Fig.~\ref{randTurb_qtrit} shows the potential advantage of the correction process.

\begin{figure}[ht]
\centering
\includegraphics[width = 0.8\linewidth]{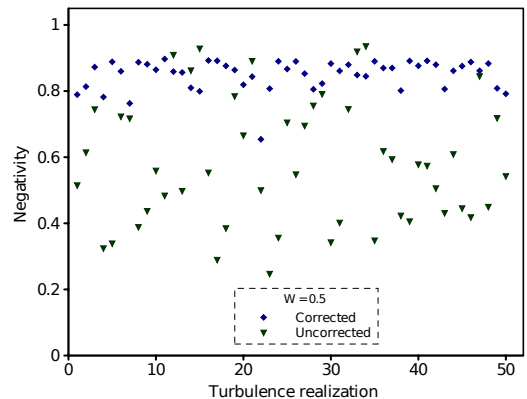}
\caption{The negativity of the correct and the uncorrected density matrices for the different atmospheric turbulence realizations. Triangle markers represent the uncorrected, truncated density matrices. Diamond markers represent the corrected density matrices.}
\label{neg_qutrit}
\end{figure}

To investigate how the entanglement (or classical non-separability) is affected by the compressive sensing correction technique, we calculate the negativity of the density matrices. The negativity of a state is given by
\begin{equation}
\mathcal{E}(\rho_\text{c}) = \frac{1}{2}\sum_{n}(|\lambda_n| - \lambda_n),
\end{equation}
where $\lambda_n$ are the eigenvalues of the partially transposed density matrix. The partial transpose of a density matrix is obtained by performing a transpose on the density matrix of one subsystem, leaving the other the same.

Figure~\ref{neg_qutrit} shows that the negativity of the state improves with application of the correction procedure that is based on the compressive sensing technique. The mean negativity for the corrected density matrices over 50 realizations is $0.83 \pm 0.04$. This value can be contrasted with that of the uncorrected, truncated density matrices, which is $0.56 \pm 0.18$, as evidence of the improvement.

\begin{figure}[ht]
\centering
\includegraphics[width = 0.8\linewidth]{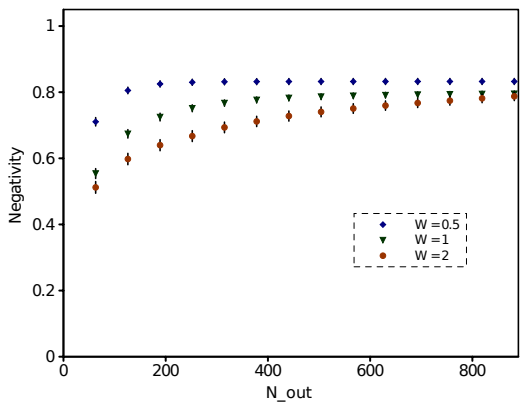}
\caption{The negativity of the corrected density matrix as a function of $N_\text{out}$ for different scintillation strengths $W$.}
\label{NegN}
\end{figure}

For the practical design of compressive sensing-based channel correction for optical and quantum communication systems, we considered the mean negativity of the corrected density matrices as a function of the output dimension $N_\text{out}$. In Fig.~\ref{NegN}, we display three curves for different scintillation strengths $W$ (by changing the propagation distance).

One can see that the mean negativity of the corrected density matrices increases with $N_\text{out}$, indicating the effect of the chosen size of the output Hilbert space. Furthermore, we observe that the mean negativity becomes saturated. The saturation level and the rate of saturation depends on the scintillation strength.

These observations suggest a possible future improvement to the proposed scheme. By applying methods such as deep learning \cite{goodfellow2016deep,nielsen2015neural} or other  generic machine learning algorithms \cite{ethem, ismail}, one may be able to train the system to determine an optimal value for $N_\text{out}$, given certain  turbulence conditions.

\section{Conclusions}
\label{conclusions}

Using a numerical analysis, we demonstrate the performance of a proposed compressive sensing-based channel correction method. A classically non-separable state, consisting of three OAM modes with different wavelengths, was used as input to a numerically simulated turbulent free-space channel. Using compressive sensing-based state tomography, we reconstructed the output state and used it to determine the Kraus operator matrix for the channel. The singular value thresholding technique was used for the compressive sensing algorithm. The results show that compressive sensing drastically reduces the number of measurements required for the characterization of the turbulent channel. Although the channel estimation process uses classical light, it determines the Kraus operator matrix for the channel and thus allows it to be used for quantum communication. Consequently, the proposed scheme would be useful in the design of quantum communication systems that are based on the OAM states of light.

As a further study, we intend to investigate the use of deep learning or machine learning as a method to inform the system on the appropriate dimension of the Kraus operator matrix. It could be done by training a model to determine an appropriate value of this dimension, based on the parameters of the turbulent free-space channel.

\section*{Acknowledgements}

CMM acknowledges support from the CSIR National Laser Centre. The research for this work was supported in part by the National Research Foundation (NRF) of South Africa (Grant Number: 118532).

\end{document}